\documentstyle[preprint,pra,aps]{revtex}

\begin{document}

\draft

\tightenlines

\title{Ultraviolet analysis of one dimensional quantum systems}

\author{Marco Frasca}
\address{Via Erasmo Gattamelata, 3,
         00176 Roma (Italy)}

\date{\today}

\maketitle

\abstract{
Starting from the study of one-dimensional potentials in quantum mechanics having
a small distance behavior described by a harmonic oscillator, we extend this way
of analysis to models where such a behavior is not generally expected. In order
to obtain significant results we approach the problem by a renormalization
group method that can give a fixed point Hamiltonian that has the shape of
a harmonic oscillator. In this way, good approximations are obtained for the
ground state both for the eigenfunction and the eigenvalue for problems like
the quartic oscillator, the one-dimensional Coulomb potential having a 
not normalizable ground state solution 
and for the one-dimensional Kramers-Henneberger potential.
We keep a coupling constant in the potential and take
it running with a generic cut-off that goes to infinity. 
The solution of the Callan-Symanzik equation
for the coupling constant generates the harmonic oscillator
Hamiltonian describing the behavior of the model at very small distances
(ultraviolet behavior). This approach, although algorithmic in its very nature,
does not appear to have a simple extension to obtain excited state behavior.
Rather, it appears as a straightforward non-perturbative method.
}


\narrowtext

\section{Introduction} 

Since its inception, renormalization group proved to be a powerful method
to study the ultraviolet behavior in quantum field theory \cite{gl}. The work
of Wilson \cite{wil} that applied a new renormalization group method to the
study of continuous phase transitions to determine the critical indices of a
given universality class has given a striking example of how powerful such
an approach can be. Recent significant results were offered by the prove
of asymptotic freedom in quantum chromodynamics \cite{wei}.

After this relevant number of successes, renormalization group method has
been applied in other fields as chaos \cite{fei} and perturbation theory \cite{gk}
giving significant improving of the possibility of analysis in such disparate fields.

The study of one dimensional problems in quantum mechanics has a relevant
importance being generally seen as the study of zero dimensional quantum field theory. From
this point of view, the pioneering analysis by Bender and Wu \cite{bw} started
an important research line that gives significant results till today. Different
methods have been devised to obtain an ever improved approximation to the exact
eigenvalues of the one dimensional anharmonic oscillator \cite{kk}. 

Besides, one dimensional problems exist in quantum mechanics where a renormalization
group approach is needed to find the ground state eigenvalue and eigenfunction
\cite{rg1}. An example is given by the one dimensional Coulomb potential $1/|x|$
where it is known to exist a ground state non-normalizable solution that is
excluded (see e.g. \cite{fai}). In this way a one dimensional electron will never
pass over the nucleus contrarily to what happens in two and three dimensions. In
this paper we will solve this problem. 

The starting point of this paper is given by the question: Does it exist a small
distance model for a given one dimensional quantum system that catches all its
properties in such a regime? This question has a simple answer when we take
a system with a Morse potential that is a straightforward generalization of
a harmonic oscillator and reduces to it in some approximation. So, we know that
a Taylor series does exists that reduces to the harmonic oscillator Hamiltonian.
Besides, the Morse potential gives an exactly solvable Schr\"odinger equation
and can be used as a fairly example of the method. The question becomes more
interesting when such a Taylor series does not appear to exist 
at small distances at a first sight.
Indeed, one can do a Taylor series around another point and in this way we can
extend the above consideration to any one dimensional system.

The renormalization group applies when we choose the point of the converging
Taylor series becoming more and more smaller (ultraviolet limit). In this case,
a renormalization of the coupling constant is needed and, looking for 
a solution of the Callan-Symanzik equation, gives rise to the 
harmonic oscillator Hamiltonian that describes the behavior of the
model at small distances. The method is inherently non-perturbative.

The paper is so structured. 
In sec.II we discuss the generalities of the method we propose
using as an example the Morse potential.
In sec.III we discuss the fundamental case of a quartic oscillator that is a
test for any approach as the one given in this paper. In sec.IV we solve the
question of the one dimensional Coulomb potential and the behavior at small
distances. In sec.V we discuss the interesting case of the Kramers-Henneberger
potential that is widely used in study of stabilization of an atom in a strong
laser field. Finally, in sec.VI the conclusions are given.

\section{A generic renormalization group approach}

The Morse potential, given by
\begin{equation}
    V(x)=A(e^{-2ax}-2e^{-ax}),
\end{equation}
is known to have as a limit case, in the ultraviolet limit $x\rightarrow 0$, the
harmonic oscillator. This means that the ground state energy and wavefunction 
can be obtained, in some approximation, by taking this limit. Indeed, in the
ultraviolet limit, one gets the ground state energy (here and in the following $\hbar=1$)
\begin{equation}
    E_0=-A+a\sqrt{\frac{A}{2m}} \label{eq:e0}
\end{equation}
being $m$ the mass of the oscillator. This is a fairly good approximation as
just a constant, independent on $A$, is missing from the true value. One may
ask if such an analysis can be extended to quantum systems that do not have
such a straightforward expansion.

Indeed, instead to consider the expansion around the point $x=0$ one can choose
another point $x=\frac{1}{\Lambda}$ being $\Lambda$ a very large, small distance
cut-off. It is easily seen that the Morse potential is now
\begin{equation}
    V(x)\approx a^2 A(\Lambda)x^2-A(\Lambda)-a^2\frac{A(\Lambda)}{\Lambda^2}
\end{equation} 
where we have supposed that the coupling constant $A$ depends on $\Lambda$. This
is a harmonic oscillator whose ground state energy is given by
\begin{equation}
   E_0(\Lambda)=a\sqrt{\frac{A(\Lambda)}{2m}}-A(\Lambda)-\frac{a^2A(\Lambda)}{\Lambda^2}
\end{equation}
and we ask this expression to not depend on the cut-off $\Lambda$
by imposing the condition 
\begin{equation}
\frac{dE_0(\Lambda)}{d\Lambda}=0. 
\end{equation}
One gets
\begin{equation}
    \frac{dA(\Lambda)}{d\Lambda}=-2\frac{\frac{a^2A(\Lambda)}{\Lambda^3}}
	{-1-\frac{a^2}{\Lambda^2}+\frac{a}{\sqrt{8mA(\Lambda)}}}
\end{equation}
and the beta function can be defined as
\begin{equation}
    \beta[A(\Lambda),\Lambda]=2\frac{a^2A(\Lambda)}
	{\Lambda^2+a^2-\frac{a\Lambda^2}{\sqrt{8mA(\Lambda)}}}.
\end{equation}
This defines the Callan-Symanzik equation for this problem and we are applying
a renormalization group method \cite{wei}. 
Notwithstanding the non-linear and involved appearance
of this equation, the solution is quite easy in the limit $\Lambda\rightarrow\infty$
being $A(\Lambda)=constant$ and the initial approach for the Morse potential
by the harmonic oscillator approximation eq.(\ref{eq:e0}) is recovered. So,
reducing to the harmonic oscillator solves the Callan-Symanzik equation in
the ultraviolet limit.

It seems that we have made an involved reformulation of the initial approximation
to the Morse potential but, as we will see, our formulation is quite general
being applicable to any case. Here we give a summary of this renormalization
group method.

Let us take the generic one dimensional Hamiltonian
\begin{equation}
    H=\frac{p^2}{2m}+gV(x)
\end{equation}
being $g$ a coupling constant,
and compute the Taylor series in the ultraviolet limit
\begin{equation}
    V(x)=V\left(\frac{1}{\Lambda}\right)+V'\left(\frac{1}{\Lambda}\right)\left(x-\frac{1}{\Lambda}\right)
	+\frac{1}{2}V''\left(\frac{1}{\Lambda}\right)\left(x-\frac{1}{\Lambda}\right)^2+
	O\left(\left(x-\frac{1}{\Lambda}\right)^3\right)
\end{equation}
being $\Lambda$ a very large cut-off. Rewrite it as
\begin{equation}
    V(x)=\frac{1}{2}V''\left(\frac{1}{\Lambda}\right)
	\left[\left(x-\frac{1}{\Lambda}+\frac{V'\left(\frac{1}{\Lambda}\right)}
	{V''\left(\frac{1}{\Lambda}\right)}\right)^2
	+2\frac{V\left(\frac{1}{\Lambda}\right)}
	{V''\left(\frac{1}{\Lambda}\right)}
	-\left(\frac{V'\left(\frac{1}{\Lambda}\right)}
	{V''\left(\frac{1}{\Lambda}\right)}\right)^2\right]
\end{equation}
Then, solve the Schr\"odinger equation with this potential, assuming the $g=g(\Lambda)$ and
obtain $E_0(\Lambda)$. Impose the condition of independence from $\Lambda$ as
\begin{equation}
    \frac{dE_0(\Lambda)}{d\Lambda}=0
\end{equation}
and get the Callan-Symanzik equation
\begin{equation}
    \frac{dg(\Lambda)}{d\ln\Lambda}=\beta[g(\Lambda),\Lambda].
\end{equation}
Solve it by reducing to the harmonic oscillator in the limit $\Lambda\rightarrow\infty$. The
solution is generally seen by reducing the system Hamiltonian 
to the harmonic oscillator Hamiltonian. This
gives the approximation for the ground state energy $E_0(\infty)$ and the approximant
wave function in the ultraviolet limit. An ambiguity can appear in the sign of the
ground state energy. We will see some examples of this procedure
in the following sections.

\section{Quartic oscillator}

A good framework to check any kind of approximate method for the evaluation of
energy eigenvalues and eigenfunctions is the quartic oscillator that has
the Hamiltonian \cite{par}
\begin{equation}
    H=p^2+x^4.
\end{equation}
We assume that a coupling constant can be introduced as
\begin{equation}
    H=p^2+gx^4
\end{equation}
and this latter Hamiltonian can be reduced to the preceding one by the scaling
$x\rightarrow g^{-\frac{1}{6}}x$ and the corresponding rescaling in the energy
eigenvalues $E\rightarrow Eg^{\frac{1}{3}}$. 
By applying the renormalization group method
here devised we obtain the Schr\"odinger equation
\begin{equation}
     -\frac{d^2\psi(x)}{dx^2}+g(\Lambda)\frac{6}{\Lambda^2}
	 \left[\left(x-\frac{2}{3\Lambda}\right)^2+\frac{1}{18\Lambda^2}\right]\psi(x)=E\psi(x)
\end{equation}  
and the energy of the ground state is
\begin{equation}
    E_0(\Lambda)=\sqrt{\frac{6g(\Lambda)}{\Lambda^2}}+\frac{g(\Lambda)}{3\Lambda^4}.
\end{equation} 
This gives rise to the beta function
\begin{equation}
    \beta[g(\Lambda),\Lambda]=2g(\Lambda)
	\frac{9\Lambda^2+2\sqrt{\frac{6g(\Lambda)}{\Lambda^2}}}
	{9\Lambda^2+\sqrt{\frac{6g(\Lambda)}{\Lambda^2}}}.
\end{equation} 
We see that the solution that turns the ultraviolet theory to
\begin{equation}
    H=p^2+x^2.
\end{equation}
is the one having
\begin{equation}
    g(\Lambda)=\frac{\Lambda^2}{6}
\end{equation}
that is, we have a continued growth of the coupling constant. This give for the ground state
energy $E_0(\infty)=1$. This is a rather good result as the exact value is around $1.06$. It is
easy to see that a simplifying procedure is to reduce in any case to the
form $p^2+x^2$ for the ultraviolet limit theory. This gives immediately the result solving the
Callan-Symanzik equation in the limit $\Lambda\rightarrow\infty$.

\section{Coulomb potential}

The one dimensional Coulomb potential has the form $1/|x|$ and characterizes un
eigenvalue problem that is not well defined having the ground state not bounded
from below and the relative eigenstate being not normalizable \cite{fai}. 
Eliminating this state as unphysical means that the electron in the ground
state cannot hit the nucleus differently from the two and three dimensional case.
Indeed, all the other eigenstates vanish at the origin. This situation is quite
similar to the case of the $1/x^2$ potential that needs a renormalization
procedure \cite{rg1}. So, in the same spirit we apply the above procedure to
the Hamiltonian
\begin{equation}
    H=\frac{p^2}{2}-\frac{\alpha}{|x|}
\end{equation}
where we have introduced the coupling $\alpha$ that can be removed by the rescaling
$x\rightarrow\frac{x}{\alpha}$ and $E\rightarrow E\alpha^2$. The Schr\"odinger
equation can be written as
\begin{equation}
     -\frac{1}{2}\frac{d^2\psi(x)}{dx^2}-\alpha(\Lambda)\Lambda^3
	 \left[\left(x-\frac{3}{2\Lambda}\right)^2+\frac{3}{4\Lambda^2}\right]\psi(x)=E\psi(x)
\end{equation} 
Then, the ground state energy is given by
\begin{equation}
    E_0(\Lambda)=\frac{1}{2}\sqrt{-2\alpha(\Lambda)\Lambda^3}-\frac{3}{4}\alpha(\Lambda)\Lambda
\end{equation}
that originates the beta function
\begin{equation}
    \beta[\alpha(\Lambda),\Lambda]=
	-3\alpha(\Lambda)
	\frac{2\Lambda+\sqrt{-2\alpha(\Lambda)\Lambda}}{2\Lambda+3\sqrt{-2\alpha(\Lambda)\Lambda}}.
\end{equation}
The solution to the Callan-Symanzik equation can be written as
\begin{equation}
    \alpha(\Lambda)=-\frac{1}{2\Lambda^3}.
\end{equation}
that gives an asymptotically free theory but coming from a direction with unstable Hamiltonians.
There is an ambiguity in the sign of the ground state energy that in this case is easily
resolved. So, finally we get $E_0(\infty)=-\frac{1}{2}$ that turns out to be the
ground state energy of the ``physical'' ground state. But now, one has a finite
probability to find the electron in the nucleus position,the origin, given by
the harmonic oscillator ground state eigenstate. Again, we see that a proper
solution of the Callan-Symanzik equation 
gives the ultraviolet limit theory $\frac{1}{2}(p^2+x^2)$ yielding
a good approximation for the ground state of the quantum system. 
So, a more direct procedure to obtain the result is to compute directly the
harmonic oscillator Hamiltonian: This gives immediately the solution of the
corresponding Callan-Symanzik equation.

Another equivalent approach can be obtained by considering the modified Coulomb
potential
\begin{equation}
    V(x)=-\frac{1}{\sqrt{x^2+\frac{1}{\Lambda^2}}}
\end{equation}
where, at the end of the computation the limit $\Lambda\rightarrow\infty$ is
taken. Now, we prove that this system has the same ultraviolet behavior of the
one dimensional Coulomb potential. Indeed, in this case the Schr\"odinger
equation takes the form
\begin{equation}
     -\frac{1}{2}\frac{d^2\psi(x)}{dx^2}-\alpha(\Lambda)\Lambda^3\frac{\sqrt{2}}{16}
	 \left[\left(x-\frac{3}{\Lambda}\right)^2+\frac{8}{\Lambda^2}\right]\psi(x)=E\psi(x)
\end{equation}
with the ground state energy
\begin{equation}
    E_0(\Lambda)=\frac{1}{2}\sqrt{-\frac{\sqrt{2}}{8}\alpha(\Lambda)\Lambda^3}
	-\frac{\sqrt{2}}{2}\alpha(\Lambda)\Lambda
\end{equation}
giving the beta function
\begin{equation}
    \beta[\alpha(\Lambda),\Lambda]=
	-\alpha(\Lambda)
	\frac{3\Lambda+4\sqrt{-2\sqrt{2}\alpha(\Lambda)\Lambda}}{\Lambda+4\sqrt{-2\sqrt{2}\alpha(\Lambda)\Lambda}}.
\end{equation} 
that proves our assertion: Both theories have the same ultraviolet behavior except
for a scale numerical factor. This result will be used in the following section.

\section{Kramers-Henneberger potential}

Kramers-Henneberger potential is a Coulomb potential dressed by the presence
of a strong monochromatic electromagnetic field \cite{pg,fra}. It is generally
considered in studies about stabilization of atom in strong laser fields. The
Schr\"odinger equation can be written for the one dimensional Coulomb
potential as
\begin{equation}
    -\frac{\hbar^2}{2m}\frac{d^2\psi(x)}{dx^2}-
	\frac{Ze^2}{\pi\lambda_L}\int_{-1}^1\frac{1}{\left|\frac{x}{\lambda_L}-x'\right|}
	\frac{dx'}{\sqrt{1-x'^2}}\psi(x)=E\psi(x) \label{eq:kh}
\end{equation}
being $\lambda_L$ the amplitude of the motion of a free particle with charge
$e$ in an electric field time varying sinusoidally in time, $Z$ the atomic
number and $m$ the mass of the particle. $\lambda_L=\frac{e{\cal E}}{m\omega^2}$
being $\cal E$ the amplitude of the electromagnetic field and $\omega$ its frequency.
Eq.(\ref{eq:kh}) is not defined as the integral as such has no meaning. A way out
is, generally, to modify the Coulomb potential in some way to remove the
singularity \cite{haa}. Here we try another approach to smooth the singularity.
We want to use the renormalization group method devised above. In order to do
this, we introduce the cut-off $\Lambda$ into the integral as 
\begin{equation}
    -\frac{\hbar^2}{2m}\frac{d^2\psi(x)}{dx^2}-
	\frac{Ze^2}{\pi\lambda_L}\int_{-1}^1
	\frac{1}{\sqrt{\left(\frac{x}{\lambda_L}-x'\right)^2+\frac{1}{\Lambda^2}}}
	\frac{dx'}{\sqrt{1-x'^2}}\psi(x)=E\psi(x) \label{eq:khm}
\end{equation}
supposing to remove the cut-off at the end of the computation taking the limit
$\Lambda\rightarrow\infty$. Rather than to consider the solution to eq.(\ref{eq:khm}), we
try an ultraviolet study of the above equation. This is the interesting limit for
a strong laser field because, in this case, one has, generally, that the amplitude
$\lambda_L$ is much greater than the Bohr radius.

Before to proceed, we change the variable by setting $z=\frac{x}{\lambda_L}$
and leave the scaled cut-off with the same name. Acting in this way, the
equation takes the scaled form
\begin{equation}
    -\frac{1}{2}\frac{d^2\psi(z)}{dz^2}-
	\frac{1}{\pi\epsilon}\int_{-1}^1
	\frac{1}{\sqrt{\left(z-z'\right)^2+\frac{1}{\Lambda^2}}}
	\frac{dz'}{\sqrt{1-z'^2}}\psi(z)=\hat{E}\psi(z)
\end{equation}
being $\epsilon=\frac{\hbar^2}{mZe^2\lambda_L}$ the ratio between the Bohr
radius and the amplitude $\lambda_L$ and $\hat{E}=E\frac{\lambda_L}{\epsilon Ze^2}$.
The Taylor series of the integral in the above equation, when $\Lambda\rightarrow\infty$,
gives $2\ln\Lambda+\ln\Lambda z^2$ where we realize that the singular behavior
of the integral is logarithmic. Then, The Schr\"odinger equation takes the
form
\begin{equation}
     -\frac{1}{2}\frac{d^2\psi(z)}{dz^2}-\frac{1}{\pi\epsilon_{exp}}\alpha(\Lambda)\ln\Lambda
	 \left[2+z^2\right]\psi(z)=\hat{E}\psi(z)
\end{equation} 
where we have introduced the experimental parameter $\epsilon_{exp}$ and set
\begin{equation}
    \alpha(\Lambda)=-\frac{\epsilon_{exp}}{\epsilon(\Lambda)}
\end{equation}
the running coupling constant. So, we have a harmonic oscillator Hamiltonian
having a ground state energy
\begin{equation}
    \hat{E}_0=\frac{1}{2}\sqrt{\frac{2}{\pi}\frac{\alpha(\Lambda)}{\epsilon_{exp}}\ln\Lambda}
	+\frac{2}{\pi}\frac{\alpha(\Lambda)}{\epsilon_{exp}}\ln\Lambda
\end{equation}
that gives the Callan-Symanzik equation in the very simple form
\begin{equation}
    \frac{d\alpha(\Lambda)}{d\ln\Lambda}=-\frac{\alpha(\Lambda)}{\ln\Lambda}
\end{equation}
that has the quite simple solution
\begin{equation}
    \alpha(\Lambda)\ln\Lambda=K^2
\end{equation}
being $K$ a constant that can depend on $\epsilon_{exp}$. The square is due to
the ambiguity in the sign of the ground state energy. We see that we have recovered
in a rigorous way, using renormalization group methods, the renormalization
method used in Ref.\cite{fra}, supporting it mathematically. Besides, we have
shown that this theory is asymptotically free.But, our aim is to
obtain an expression, even if approximated, of the ground state energy that, in
three dimensions, is known just numerically \cite{pg}. The only criteria we have
to get the ground state energy is that, for $\epsilon_{exp}$ going to zero, we
have to recover the well-known result for hydrogen atom already seen in the
preceding section. We do not care about the exact form of $K$ as a function
of $\epsilon_{exp}$ but we note that when $\epsilon_{exp}\rightarrow\infty$ 
(small fields) the proper choice is $K=-\sqrt{\frac{2}{\pi\epsilon_{exp}^3}}$
that gives
\begin{equation}
    E_0\approx-\frac{\cal R}{2}+\frac{\cal R}{\epsilon_{exp}^2}
\end{equation}
being $\cal R$ the Rydberg constant,
to be compared with the three dimensional case given in Ref.\cite{pg} as
\begin{equation}
    E_0\approx-\frac{\cal R}{2}+\frac{\cal R}{3\epsilon_{exp}^2}
\end{equation}
accounting for the Stark shift. When $\epsilon_{exp}\rightarrow 0$ the proper
choice is $K^2=\epsilon_{exp}$ giving
\begin{equation}
    E_0\approx\pm\frac{\cal R}{2}\sqrt{\frac{2}{\pi}}\epsilon_{exp}^2+\frac{2}{\pi}{\cal R}\epsilon_{exp}^2
\end{equation}
completing the agreement with the renormalization method of Ref.\cite{fra}
even if we are in trouble with the sign ambiguity. In any case we have
$E_0\propto{\cal R}\epsilon_{exp}^2$ and greater than zero. The state is anyhow
localized as we are representing the system by a harmonic oscillator in the ground state.
It is interesting to note that the zone where the particle is localized is given by
the length $\lambda_L$ and this agrees with the model generally accepted for the
behavior of an electron in a strong field \cite{cork}. In this way we can say
that the atom indeed stabilizes against the ionizing effect of the electromagnetic
field.
 
\section{Conclusions}

We have described an approach based on the renormalization group to determine
the ultraviolet behavior of one dimensional quantum systems. The ultraviolet
limit theory is given by the quantum harmonic oscillator that, in turn, gives
an approximation to the ground state energy and wave function of the quantum
system we aim to describe. The method gives finite results also for models
whose ground state wave function is not normalizable as the one dimensional
Coulomb potential and permits to obtain meaningful results through a proper
redefinition of the quantum problem as for the Kramers-Henneberger potential.

Notwithstanding the interesting results for the simple models we have
considered, there can be situation where the ultraviolet limit theory
does not exist. Besides, we have not been able to find a proper extension to
the determination of the behavior of excited states. Finally, there is an
ambiguity in the sign to determine the ground state energy that, sometimes,
it is not possible to resolve. Anyhow, this approach can offer an alternative
methods to perturbation theory and can give some significant indication of
the behavior of a quantum system at small distances. 

\label{end}

\end{document}